\documentstyle[12pt]{article}
\begin{document}
\def \beq{\begin{equation}}
\def \eeq{\end{equation}}
\def \bea{\begin{eqnarray}}
\def \eea{\end{eqnarray}}
\def \tt {t\overline {t}}
\def \ee {e^+e^-}
\def \ra {\rightarrow}
\def \g {\gamma}
\def \cvg {c_v^{\gamma}}
\def \cag {c_a^{\gamma}}
\def \cvz {c_v^Z}
\def \caz {c_a^Z}
\def \cdg {c_d^{\gamma}}
\def \cdz {c_d^Z}
\def \cdgz {c_d^{\gamma,Z}}
\def \t{$t\;$}
\def \toverline{$\overline t$}
\def \tbar{$\overline t\;$}
\def \el{E_l}
\def \thetal{\theta_l}
\def \phil{\phi_l}
\def \CP{$CP$}
\def \f{\frac}
\def \o{\overline}
\def \n{\noindent}
\begin{flushright}
{\large FTUV/96-36 IFIC/96-44}\\
{\large hep-ph/9606356}
\end{flushright}
\vskip .25cm
\begin{center}
{\huge Simple decay-lepton asymmetries in polarized $e^+e^- \ra
\tt$ and $CP$-violating dipole couplings of the top quark }
\vskip .5cm
{\large P. Poulose}\\
{\it Theory Group, Physical Research Laboratory \\ Navrangpura,
Ahmedabad 380009, India}
\vskip .3cm
{\large Saurabh D. Rindani\footnote{On sabbatical leave 
from Theory Group, Physical Research Laboratory, Navrangpura,
Ahmedabad 380 009, India, until September 19, 1996}}\\
{\it Institut de F\' \i sica Corpuscular\\ 
Departament de F\'\i sica Te\` orica , 
Universitat de Val\` encia\\ 
Av. Dr. Moliner 50, 
46100 Burjassot, Val\` encia, 
Spain
}
\vskip 1cm
{\bf Abstract}
\end{center}
\vskip .2cm
We study two 
simple $CP$-violating asymmetries of leptons coming from the decay of
$t$ and $\o t$ in $e^+e^- \ra \tt$, which do not need the
full reconstruction of the $t$ or $\o t$ for their measurement.
They can arise when the top  quark
possesses nonzero electric and weak dipole form factors in the couplings to
the photon and $Z$, respectively. Together, these two asymmetries can help to
determine the electric and weak dipole form factors independently. 
If longitudinal beam polarization is available, independent determination
of form factors can be done by measuring only one of the asymmetries. 
We obtain estimates of 90\% confidence limits that can be put on these
form factors at a future linear $\ee$ collider operating at $\sqrt{s}=500$ GeV.
\newpage

Experiments at the Tevatron have seen evidence for the top qark with
mass in  
the range of about 170-200 GeV \cite{expt}.  Future runs of the
experiment will be able to  
determine the mass more precisely and also determine other properties
of the top quark.  $\tt$ pairs will be produced more copiously at
proposed 
$e^+ e^-$ linear colliders operating above threshold.  It would then
be  
possible to investigate these properties further. 

While the standard model (SM) predicts $CP$ violation outside
the \mbox{$K$-,} 
$D$- and $B$-meson systems to be unobservably small, 
in some extensions of SM, $CP$ violation might be considerably 
enhanced, especially in the presence of a heavy top quark.  In
particular, 
$CP$-violating electric dipole form factor of the top quark, and 
the analogous $CP$-violating ``weak" dipole form factor in the $\tt$
coupling 
to $Z$, could be enhanced.  These $CP$-violating form factors could
be  
determined in a model-independent way at high energy $\ee$ linear
colliders, 
where $\ee \ra \tt$ would proceed through virtual $\g$ and $Z$
exchange. 

Since a heavy top quark ($m_t \ge 120$ GeV) is expected to decay
before it  
hadronizes \cite{heavytop}, it has been suggested \cite{toppol} that
top 
polarization asymmetry in   
$\ee \ra \tt$ can be used to determine the $CP$-violating dipole form
factors, 
since polarization information would be retained in the decay product
distribution.  Experiments have been proposed in which the
$CP$-violating 
dipole couplings could be measured in decay momentum
correlations$^1$ \cite{bern,atwood,cuypers} or 
asymmetries \cite{asymm,PP}, even with beam polarization
\cite{atwood,PP}.  These suggestions on the  
measurement of asymmetries have concentrated on experiments requiring
the reconstruction of the top-quark momentum (with the exception of
lepton  energy asymmetry \cite{toppol,asymm,PP}).  In this note we 
look at very simple lepton angular asymmetries 
in $e^+e^- \ra \tt$ 
which do not require the experimental determination of the $t$ or
$\o{t}$ momentum.  Being single-lepton asymmetries, they do not
require  
both $t$ and $\o{t}$ to decay leptonically. Since either $t$ or
$\o{t}$ 
is also allowed to decay hadronically, there is a gain in statistics.

The two asymmetries we study here are as follows. We look at the angular
distributions of the charged leptons arising from the decay of $t$ and $\o t$
in $e^+e^- \ra \tt$. In terms of the polar angle distribution of the leptons
with respect to the $e^-$ beam direction in the centre-of-mass (cm) frame, we
can define two $CP$-violating asymmetries.
One is simply the 
total lepton-charge asymmetry, with a cut-off $\theta_0$ on the forward
and backward polar angles of the leptons, with respect to the beam
direction as $z$ axis. The other is the leptonic forward-backward asymmetry
combined with charge asymmetry, again with the angles within 
$\theta_0$ of the forward and
backward directions excluded. (See later for details).

Our results are based on a fully analytical calculation of single
lepton  
distributions in the production and subsequent decay of $\tt$. We
present here only the leptonic asymmetries obtained by an integration
of these distributions. The details of the 
fully differential distribution as well as the distribution in the  
polar angle of the lepton with respect to the beam direction in the  
centre-of-mass (cm) frame 
can be found elsewhere \cite{longpaper}.

We have also included the effect of electron longitudinal
polarization, 
likely to be easily available at linear colliders.  In an
earlier paper \cite{PP},
we had shown how polarization helps to put independent limits on
electric  
and weak dipole couplings, while providing greater sensitivity in the
case of asymmetries.  We also demonstrate these 
advantages for the present case, strengthening the case for
polarization studies.

We first describe the calculation of these asymmetries in terms of 
the electric and weak dipole couplings of the top quarks. These $CP$-violating
couplings give rise to top polarization asymmetries in the production of 
$\tt$ in  
$\ee \ra \tt$ which in turn give rise to angular asymmetries in the subsequent decay $t \ra b l^+ \nu_l\;
(\overline{t} \ra 
\overline{b} l^- \overline{\nu_l})$.  We adopt the narrow-width
approximation for  
$t$ and $\overline{t}$, as well as for $W^{\pm}$ produced in
$t,\;\overline{t}$ 
decay.

  We assume the top quark couplings to $\g$ and $Z$ to be
given by the vertex factor  $ie\Gamma_\mu^j$, where
\beq
\Gamma_\mu^j\;=\;c_v^j\,\g_\mu\;+\;c_a^j\,\g_\mu\,\g_5\;+
\;\f{c_d^j}{2\,m_t}\,i\g_5\,
(p_t\,-\,p_{\overline{t}})_{\mu},\;\;j\;=\;\g,Z, 
\eeq
with
\bea
\cvg&=&\f{2}{3},\:\;\;\cag\;=\;0, \nonumber \\
\cdz&=&\f {\left(\f{1}{4}-\f{2}{3} \,x_w\right)}
{\sqrt{x_w\,(1-x_w)}},  
 \\
\caz&=&-\f{1}{4\sqrt{x_w\,(1-x_w)}}, \nonumber
\eea
and $x_w=sin^2\theta_w$, $\theta_w$ being the weak mixing
angle.
We have assumed in (1) that the only addition to the SM
couplings $c^{{\g},Z}_{v,a}$ are the $CP$-violating electric and weak
dipole 
form factors, $e\cdg/m_t$ and $e\cdz/m_t$, which are
assumed small.  Use has also been made of the Dirac equation in
rewriting the usual dipole coupling
$\sigma_{\mu\nu}(p_t+p_{\overline{t}})^{\nu}\g_5$ as
$i\g_5(p_t-p_{\overline{t}})_{\mu}$, dropping small corrections
to the vector and axial-vector couplings.  We assume that there
is no $CP$ violation in $t$, $\overline{t}$ decay$^2$.

The helicity amplitudes for $\ee \ra \g^*,Z^* \ra \tt$ in the
cm frame, including $\cdgz$ couplings, have been
given in \cite{asymm} (see  
also Kane {\it et al.}, ref. \cite{toppol}).
We have calculated the 
helicity amplitudes for 
\[t \ra b W^+,\;
\;W^+ \ra l^+ \nu_l\]
and\[\overline{t} \ra \overline{b}W^-,\;\; W^- \ra
l^-\overline{\nu_l}\] 
in the respective rest frames of $t$, $\overline{t}$, 
assuming standard model couplings and neglecting all masses
except $m_t$, the top mass. The expressions for these can be found in 
\cite{longpaper}.

Combining the production and decay amplitudes in the narrow-width 
approximation for $t,\overline{t},W^+,W^-$, and using appropriate
Lorentz  
boosts to calculate everything in the $\ee$ cm frame, we obtained the
$l^+$  
and $l^-$ distributions for the case of $e^-$, $e^+$ with
polarization  
$P_e$, $P_{\o e}$, the expressions can again be found in \cite{longpaper}. 
We further carry out the necessary integrations to obtain only the 
polar angle distributions for the leptons, which we use to write down 
the expressions for the $CP$-violating asymmetries defined below.

We define two independent \CP-violating asymmetries, which depend on
different  
linear combinations of Im$\cdg$ and Im$\cdz$ . (It is not
possible to define \CP-odd 
 quantities which determine Re$\cdgz$ using single-lepton
distributions \cite{longpaper}).
One is simply the total  
lepton-charge asymmetry, with a cut-off of $\theta_0$ on the forward
and backward directions:
\beq
{\cal A}_{ch}(\theta_0)=\frac{
{\displaystyle		\int_{\theta_0}^{\pi-\theta_0}}d\theta_l
{\displaystyle          \left( \frac{d\sigma^+}{d\theta_l}
		-	\frac{d\sigma^-}{d\theta_l}\right)}}
{
{\displaystyle		\int_{\theta_0}^{\pi-\theta_0}}d\theta_l
{\displaystyle          \left( \frac{d\sigma^+}{d\theta_l} +
\frac{d\sigma^-}{d\theta_l}\right)}}.
\eeq
The other is the lept\-onic forward-backward asy\-mmetry com\-bined
with charge asy\-mmetry, again with the angles within $\theta_0$ of
the forward and back\-ward directions excluded:
\beq
{\cal A}_{fb}(\theta_0)= \frac{ {\displaystyle
\int_{\theta_0}^{\frac{\pi}{2}}}d\theta_l {\displaystyle
\left( \frac{d\sigma^+}{d\theta_l} -
\frac{d\sigma^-}{d\theta_l}\right)} {\displaystyle
-\int^{\pi-\theta_0}_{\frac{\pi}{2}}}d\theta_l {\displaystyle
\left( \frac{d\sigma^+}{d\theta_l} -	\frac{d\sigma^-}{d\theta_l}
\right)}}
{
{\displaystyle		\int_{\theta_0}^{\pi-\theta_0}}d\theta_l
{\displaystyle          \left( \frac{d\sigma^+}{d\theta_l} +
\frac{d\sigma^-}{d\theta_l}\right)}}.
\eeq

In the above equations,
$\sigma^+$ and $\sigma^-$ refer respectively to the cm $l^+$ and
$l^-$ distributions.
$\thetal$ is used to represent the polar
angle angle of either $l^+$ or $l^-$, 
with the $z$ axis chosen
along the $e^-$ momentum.

These asymmetries are a measure of $CP$ violation in the unpolarized
case and in the case when polarization is present, but
$P_e=-P_{\overline{e}}$.  When $P_e\neq -P_{\overline{e}}$, the
initial state is not invariant under $CP$, and therefore
$CP$-invariant interactions can contribute to the asymmetries.
However, to the leading order in $\alpha$, these $CP$-invariant
contributions vanish in the limit $m_e=0$.  Order-$\alpha$ collinear
helicity-flip photon emission can give a $CP$-even contribution.
However, this background can be suppressed by a suitable cut on the
visible energy.

The expressions for ${\cal A}_{ch}(\theta_0)$ and ${\cal
A}_{fb}(\theta_0)$ 
and are given below.

\begin{eqnarray}
\lefteqn{{\cal A}_{ch}(\theta_0)=\frac{1}{2\sigma
(\theta_0)}\frac{3\pi\alpha^2} {4s}B_tB_{\o
t}\,2\cos\theta_0\sin^2\theta_0
\left(
(1-\beta^2)\log\frac{1+\beta}{1-\beta}-2\beta\right)}\nonumber\\
&&\times\left( {\rm Im}\cdg \left\{\left[ 2 c_v^
{\gamma}+(r_L+r_R)c_v^Z\right](1-P_eP_{\overline
e})+(r_L-r_R)c_v^Z(P_{\overline e}-P_e)
\right\} \right.\nonumber \\&&\left.
+ {\rm Im}\cdz \left\{\left[ (r_L+r_R) c_v^
{\gamma}+(r_L^2+r_R^2)c_v^Z\right](1-P_eP_{\overline e})+
\left[(r_L-r_R)c_v^{\gamma}\right.\right.\right.\nonumber\\
&&\left.\left.\left.+(r_L^2-r_R^2)c_v^Z\right] (P_{\overline
e}-P_e)\right\} \right);
\end{eqnarray}
\begin{eqnarray}
{\cal A}_{fb}(\theta_0)&=&\frac{1}{2\sigma
(\theta_0)}\frac{3\pi\alpha^2} {2s}B_tB_{\o t}\,\cos^2\theta_0
\left(
(1-\beta^2)\log\frac{1+\beta}{1-\beta}-2\beta\right)c_a^Z\nonumber\\
&&\times \left\{ {\rm Im}\cdg \left[ (r_L-r_R)(1-P_eP_{\overline
e})+(r_L+r_R)(P_{\overline e}-P_e)
\right] \right.\nonumber \\&&\left.
+ {\rm Im}\cdz \left[ (r_L^2-r_R^2)(1-P_eP_{\overline e})
+(r_L^2+r_R^2)(P_{\overline e}-P_e)\right]\right\}.
\end{eqnarray}
Here $\sigma(\theta_0)$ is the cross section for $l^+$ or $l^-$
production with a cut-off $\theta_0$, and is given by
\begin{eqnarray}
\sigma(\theta_0)&=& \frac{3\pi\alpha^2}{8s}B_tB_{\o t} \,
2\cos\theta_0\left(
\left\{(1-\beta^2)\log\frac{1+\beta}{1-\beta}\sin^2\theta_0
\right. \right. \nonumber \\
&& \left. \left. + 2\beta\left[1+(1-\frac{2}{3}
\beta^2)\cos^2\theta_0\right]\right\} \right.\nonumber\\ 
&& \times \left.
\left\{\left[2{c_v^{\g}}^2+2c_v^{\g}c_v^Z(r_L+r_R)+{c_v^Z}^2
(r_L^2+r_R^2)\right](1-P_eP_{\overline e})\right.\right.\nonumber \\
&&\left. \left. +c_v^Z\left[(r_L-r_R)c_v^{\g} +
(r_L^2-r_R^2)c_v^Z\right](P_{\overline e}-P_e)\right\}\right.
\nonumber \\
&&+\left.\left\{(1-\beta^2)\log\frac{1+\beta}{1-\beta}\sin^2\theta_0
+ 2\beta\left[2\beta^2-1+(1-\frac{2}{3}
\beta^2)\cos^2\theta_0\right]\right\}\right.\nonumber \\
&&\times\left. {c_a^Z}^2 \left\{(r_L^2+r_R^2)(1-P_eP_{\overline e}) +
(r_L^2-r_R^2) (P_{\overline e}-P_e)\right\} -2(1-\beta^2)\right.
\nonumber \\ &&\left. \times\left( \log\frac{ 1+\beta}{1-\beta} - 2
\right)\sin^2 \theta_0
c_a^Z\left\{\left[(r_L+r_R)c_v^{\g} + (r_L^2+r_R^2) c_v^Z \right]
\right.\right. \nonumber \\ &&\times\left. \left.(1-P_eP_{\overline
e}) + \left[ (r_L-r_R)c_v^{\g}+ (r_L^2-r_R^2) c_v^Z
\right] (P_{\overline e}-P_e)\right\}
\right).
\end{eqnarray}

In these equations, $\beta$
is the $t$ (or $\o t$) velocity: \(\beta=\sqrt{1-4m_t^2/s}\), and
$\gamma = 1/\sqrt{1-\beta^2}$, and 
$B_t$ and $B_{\o t}$ are respectively the branching ratios of
$t$ and
$\o t$ into the final states being considered.
$-er_{L,R}/s$ is the product of the $Z$-propagator and
left-handed (right-handed) electron couplings to $Z$, with
\bea
r_L&=&\f{\left(\f{1}{2}-x_w\right)}{\left(1-\f{m_Z^2}{s}\right)
\,\sqrt{x_w\,(1-x_w)}},
\nonumber \\
r_R&=&\f{-x_w}{\left(1-\f{m_Z^2}{s}\right)\,\sqrt{x_w\,(1-x_w)}}.
\eea

We note the curious fact that ${\cal A}_{ch}(\theta_0)$ vanishes for
$\theta_0=0$. This implies that the $CP$-violating charge asymmetry
does not exist unless a cut-off is imposed on the lepton production
angle. ${\cal A}_{fb}(\theta_0)$, however, is nonzero for
$\theta_0=0$.

We now describe the numerical results for the calculation
of 90\% confidence level (CL) limits that could be put on Im$\cdgz$
using the asymmetries described earlier, as well as
the $CP$-odd part of the angular distribution in eq. (9).

We look at only semileptonic final states. That is to say, when $t$
decays leptonically, we assume $\o t$ decays hadronically, and {\it
vice versa}. We sum over the electron and muon decay channels. Thus,
$B_tB_{\o t}$ is taken to be $2/3\times2/9$.  The number of events
for various relevant $\theta_0$ and for beam polarizations $P_e=0$,
$\pm 0.5$ are listed in Table 1.

In each case we have derived simultaneous 90\% CL limits on
Im$c_d^{\g}$ and Im$c_d^Z$ that could be put in an experiment at a
future linear collider with $\sqrt{s}=500$ GeV and an integrated
luminosity of 10 fb$^{-1}$. We do this by equating the asymmetry
(${\cal A}_{ch}$ or ${\cal A}_{fb}$) to $2.15/\sqrt{N}$, where $N$ is
the total number of expected events. In the unpolarized case, each of
${\cal A}_{ch}$ and ${\cal A}_{fb}$ gives a band of allowed values in
the Im$c_d^{\g}-$Im$c_d^Z$ plane. If both ${\cal A}_{ch}$ and ${\cal
A}_{fb}$ are looked for in an experiment, the intersection region of
the corresponding bands determines the best 90\% CL limits which can
be put simultaneously on Im$c_d^{\g}$ and Im$c_d^Z$. These best
results are obtained for $\theta_0=35^\circ$ and are shown in Fig.
1(a) and Fig. 1(b), for two values of the top mass, $m_t=174$ GeV,
and $m_t=200$ GeV respectively.

We see  from Fig. 1 that the 90\% CL limits that could be put on
Im$\cdg$ and Im$\cdz$ simultaneously are, respectively, 2.4 and 17,
for $m_t=174$ GeV. The same limits are 4.0 and 28 for $m_t=200$ GeV.

In the case where the $e^-$ beam is longitudinally polarized, we have
assumed the degree of polarization $P_e=\pm 0.5$, and determined 90\%
CL limits which can be achieved. In this case, the use of $P_e=+0.5$
and $P_e=-0.5$ is sufficient to constrain Im$c_d^{\g}$ and Im$c_d^Z$
simultaneously even though only one asymmetry (either ${\cal A}_{ch}$
or ${\cal A}_{fb}$) is determined. The 90\% CL bands corresponding to
$P_e=\pm0.5$ are shown in Figs. 2 and 3, for ${\cal A}_{ch}$ with
$\theta_0=60^\circ$, and for ${\cal A}_{fb}$ with
$\theta_0=10^\circ$, respectively. Again, these values of $\theta_0$
are chosen to maximize the sensitivity$^3$.

It can be seen from these figures that the simultaneous limits
expected to be obtained on Im$\cdg$ and Im$\cdz$ are, respectively,
about 0.45 and 1.5 for $m_t=174$ GeV from both the types of
asymmetries.  These limits are about 0.78 and 2.5 for $m_t=200$ GeV.
We see thus that the use of polarization leads to an improvement of
by a factor of about 5 in the sensitivity to the measurement of
Im$\cdg$, and by a factor of at least 10 in the case of Im$\cdz$.
Moreover, with polarization, either of ${\cal A}_{fb}$ and ${\cal
A}_{ch}$, with a suitably chosen cut-off, suffices to get the same
improvement in sensitivity.

Apart from simultaneous limits on Im$\cdgz$, we have also found out
the sensitivities of one of Im$\cdgz$, assuming the other to be zero,
using the $CP$-odd combination of angular distributions
$\f{d\sigma^+}{d\cos\theta} (\theta_l) - \f{d\sigma^-}{d\cos\theta}
(\pi-\theta_l)$. We assume that the data is
collected over bins in $\theta_l$, and add the 90\% CL limits
obtained from individual bins in inverse quadrature. We find that the
best individual limits are respectively 0.12 and 0.28 for Im$\cdg$
and Im$\cdz$, both in the case of $P_e=-0.5$, for $m_t=174$ GeV. The
corresponding limits for $m_t=200$ GeV are 0.18 and 0.43. As
expected, these limits are better than simultaneous ones. Even here,
there is an improvement due to polarization, but it is not as
dramatic as in the case of simultaneous limits.

Our limits on Im$\cdgz$ are summarized in Table 2.

To conclude, we have
obtained expressions for certain simple $CP$-violating angular
asymmetries
in the production and subsequent decay of $\tt$ in the
presence of electric and weak dipole form factors of the top quark.
These asymmetries are specially chosen so that they do not require the
reconstruction of the  $t$ or $\o t$ directions or energies. 
We have also included the effect of longitudinal electron beam polarization.
We have
analyzed these asymmetries to obtain simultaneous 90\% CL limits on
the imaginary parts of the electric and weak dipole couplings which
would be possible at future linear $\ee$ collider operating at
$\sqrt{s}= 500$ GeV and with a luminosity of 10 fb$^{-1}$. Figs. 1-3
show the allowed regions in the Im$\cdg$--Im$\cdz$ plane at the 90\%
CL. Table 2 summarizes the 90\% CL limits on Im$\cdgz$ in various
cases.

Our general conclusion is that the sensitivity to the measurement of
dipole couplings is improved considerably if the electron beam is
polarized, a situation which might easily obtain at linear colliders.
Another general observation is that the sensitivity is better for a
lower top mass than a higher one.

If we compare these results for sensitivities with those obtained in
\cite{PP}, where we studied asymmetries requiring the top momentum
determination, we find that while the sensitivities with the
asymmetries studied here are worse by a factor of about 3 in the
unpolarized case, the limits in the polarized case are higher by a
factor of about 2 as compared to those in \cite{PP}. It is likely
that since in the experiments suggested here, only the lepton charges
and direction 
need be determined, improvement in  experimental accuracy can easily
compensate for these factors. A detailed simulation of experimental
conditions is needed to reach a definite conclusion on the exact
overall sensitivities.

We have also compared our results with those of \cite{cuypers}, where
CP-odd momentum correlations are studied in the presence of $e^-$
polarization.  With comparable parameters, the sensitivities we
obtain are comparable to those obtained in \cite{cuypers}.  In some
cases our sensitivities are slightly worse because we require either
$t$ or $\overline{t}$ to decay leptonically, leading to a reduced
event rate.  However, the better experimental efficiencies in lepton
momentum measurement may again compensate for this loss.

As mentioned earlier, since we consider only the electron beam to be
polarized, the asymmetries considered here can have backgrounds from
order-$\alpha$ collinear initial-state photon emission, which, in
principle,  have to be calculated and subtracted. However, in case of
correlations, it was found in \cite{back} that the background
contribution can be neglected for the luminosity we assume here. This
is likely to be the case in the asymmetries we consider here.

The theoretical predictions for $c_d^{\g,Z}$ are at the level of
$10^{-2}-10^{-3} $, as for example, in the Higgs-exchange and
supersymmetric models of CP violation \cite{bern,asymm,new}.  Hence
the measurements suggested here cannot exclude these modes at the
90\% C.L.  However, as simultaneous model-independent limits on both
$c_d^{Z}$ and $c_d^{\g}$, the ones obtainable from the experiments we
suggest, are an improvement over those obtainable from measurements
in unpolarized experiments.

Increase in polarization beyond $\pm 0.5$ can increase the
asymmetries in some cases we consider.  Also, a change in the
$e^+\,e^-$ cm energy also has an effect on the asymmetries.  However,
we have tried to give here only the salient features  of the outcome
of a possible experiment in the presence of longitudinal beam
polarization.

It is obvious that the success of our proposal depends crucially on 
proper identification of the $\tt$ events and measurements of the polar
angles and the charges of leptons. This will require cuts, and will lead
to experimetnal detection efficiencies less than one as assumed here.
Our results are quantitatively exact under ideal experimental conditions.
Inclusion of experimental detection efficiencies may change our
results somewhat.  However, the main thrust of our conclusions, that
we have identified rather simple observables for measurement of dipole 
form factors, and that longitudinal beam polarization improves the 
sensitivity, would still remain valid.

\newpage
\centerline{\Large \bf Footnotes}
\vskip 1.5cm
\begin{enumerate}
\item The paper of Atwood and Soni \cite{atwood} 
introduces optimal variables whose
expectation values maximize the statistical sensitivity.
\item CP violation in top decays has  been
considered, for example, in \cite{decay}.
\item In case of ${\cal A}_{fb}$, the choice
$\theta_0=0$ also gives very similar sensitivities. However, since
this condition would be impossible to achieve in a practical
situation, we choose a nonzero value of $\theta_0$.
\end{enumerate}

\newpage
\def \n {\noindent}
\def \v{\vskip .5cm}
\centerline{\Large \bf Figure Captions}
\vskip 1.5cm
\n {\bf Fig. 1.} Bands showing simultaneous 90\% CL limits on 
Im $c_d^{\gamma}$ and
Im $c_d^Z$ using ${\cal A}_{fb}$ and ${\cal A}_{ch}$ with unpolarized
electron beam at cm energy 500 GeV and cut-off angle $35^{\circ}$.
Mass of the top quark is taken to be (a) 174 GeV and (b) 200 GeV.
\v

\n {\bf Fig. 2.} Bands showing simultaneous 90\% CL limits on 
Im $c_d^{\gamma}$ and
Im $c_d^Z$ using ${\cal A}_{ch}$ with different beam polarizations,
and at a cm energy of 500 GeV and cut-off angle $60^{\circ}$.  Mass
of the top quark is taken to be (a) 174 GeV and (b) 200 GeV.
\v

\n {\bf Fig. 3.} Bands showing simultaneous 90\% CL limits on 
Im $c_d^{\gamma}$ and
Im $c_d^Z$ using ${\cal A}_{fb}$ with different beam polarizations,
and at a cm energy of 500 GeV and cut-off angle $10^{\circ}$. Mass
of the top quark is taken to be (a) 174 GeV and (b) 200 GeV.

\def \v{\vskip .5cm}
\newpage
\v
\centerline{\Large \bf Table Captions}
\vskip 1.5cm
\n {\bf Table 1.} Number of $t \bar t$ events, with either \t  or
\tbar   
decaying leptonically,   
for c.m. energy 500 GeV and integrated luminosity $10\;{\rm fb}^{-1}$
for two different top masses with polarized and unpolarized electron
beams at different cut-off angles $\theta_0$.
\v
\n{\bf Table 2.}  Limits on dipole couplings obtainable from different
asymmetries. In case (a) limits are obtained from ${\cal A}_{ch}$ and
${\cal A}_{fb}$ using unpolarized beams (Fig. 1), and in case (b)
from either of ${\cal A}_{ch}$ (Fig. 2) and ${\cal A}_{fb}$ (Fig. 3)
with polarizations $P_e=0,\;\pm 0.5$.  Charge-asymmetric angular
distribution is used in case (c) where 0 and $\pm 0.5$
polarizations are considered separately.  All the limits are at 90\%
CL.

\newpage
\begin{center}
\begin{tabular}{||c|c|c|c||c|c|c||}
\hline
&\multicolumn{3}{c||}{$m_t=174$ GeV}&\multicolumn{3}{c||}{$m_t=200$
GeV}\\
$\theta_0$& $P_e=-0.5$&
$P_e=~0$&$P_e=+0.5$&$P_e=-0.5$&$P_e=~0$&$P_e=+0.5$ 
\\
\hline
$0^\circ$&1003&845&687 &862&723&585\\
$10^\circ$&988&832&675 &849&712&576\\
$35^\circ$&826&689&553 &711&593&475\\
$60^\circ$&507&419&330 &438&362&286\\
\hline
\multicolumn{7}{c}{}\\
\multicolumn{7}{c}{Table 1}\\
\multicolumn{7}{c}{}
\end{tabular}
\vskip 2in
\begin{tabular}{|ll|c|c|c|c|}
\hline
\multicolumn{2}{|l|}{}&\multicolumn{2}{c}{$m_t=174$ GeV}&
\multicolumn{2}{c|}{$m_t=200$ GeV}\\[3mm]
\multicolumn{2}{|l|}{Case}&$|{\rm Im}c_d^{\gamma}|$&$|$Im$
c_d^Z|$&$|$Im$c_d^{\gamma}|$&$|$Im$c_d^Z|$\\[2mm]
\hline
(a)  unpolarized&&2.4\hskip 5mm&17&  4.0&28\\[3mm]
(b)  polarized($P_e= 0,\;\pm$0.5)&&0.45&1.5&0.78&2.5\\[3mm]
(c)  angular distribution:&$P_e=+0.5$&0.13&0.74&0.21&1.21 \\
&$P_e=~~0.0$&0.13&0.81&0.20&1.30\\
&$P_e=-0.5$&0.12&0.28&0.18&0.43\\[2mm]
\hline
\multicolumn{6}{c}{}\\
\multicolumn{6}{c}{Table 2}\\
\multicolumn{6}{c}{}
\end{tabular}
\end{center}
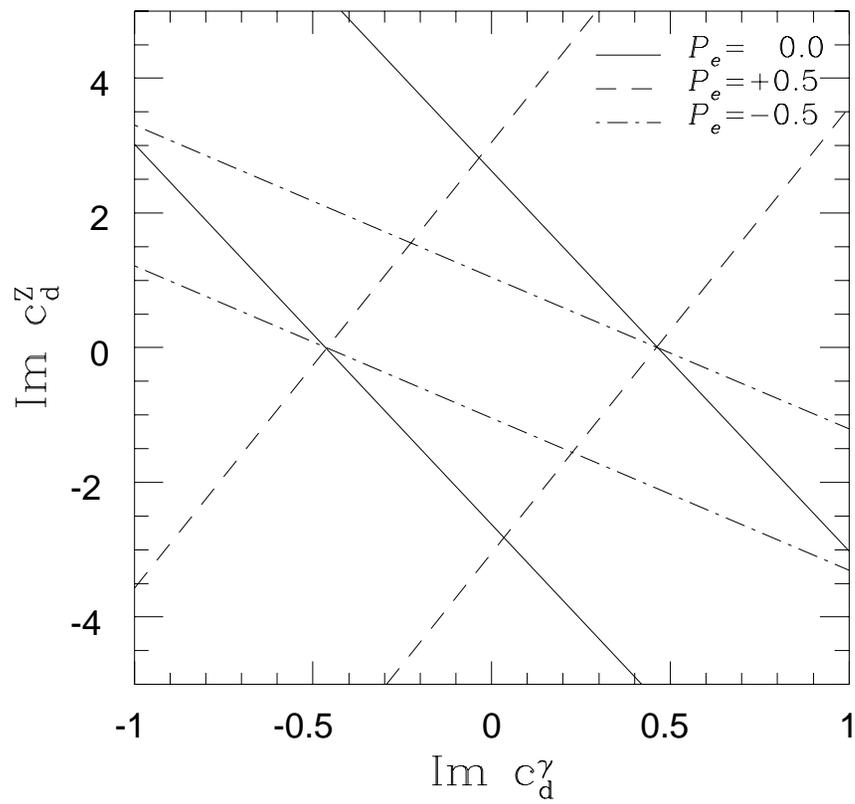
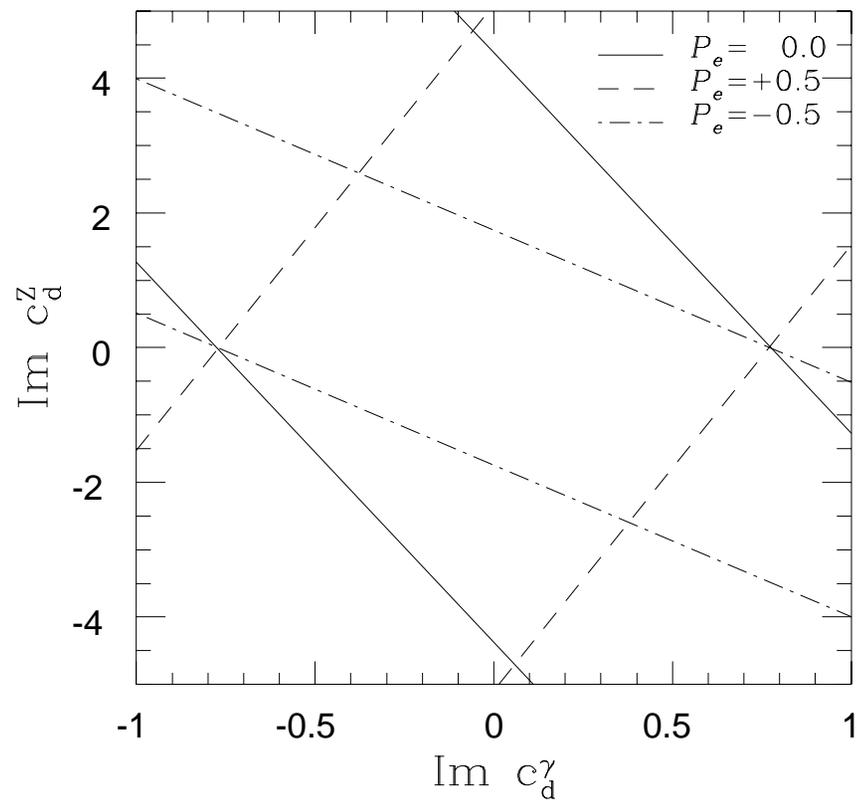
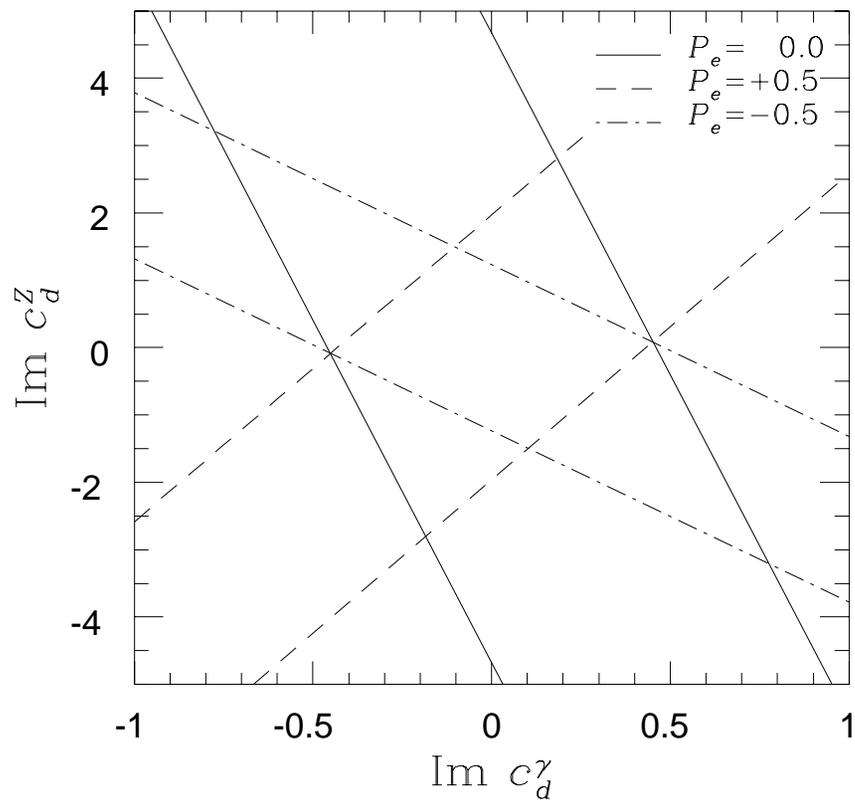
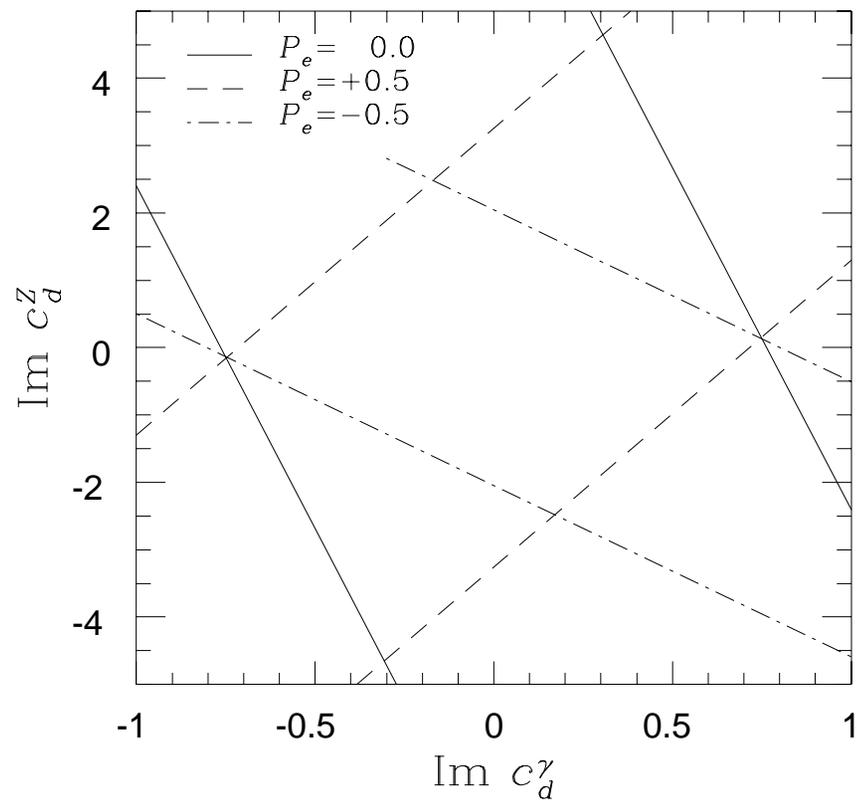
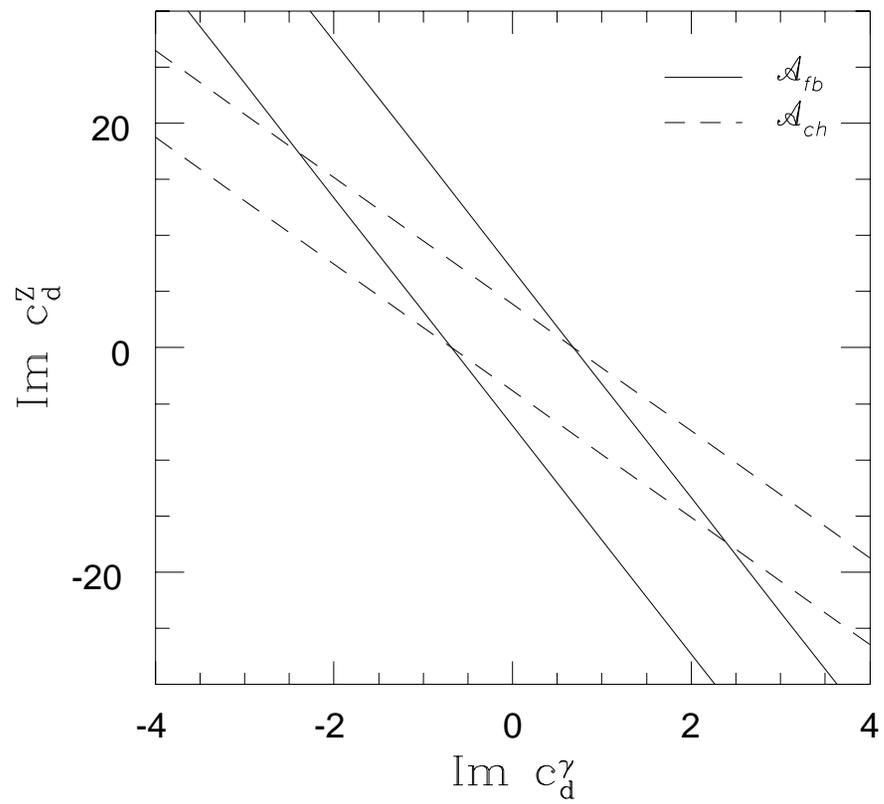
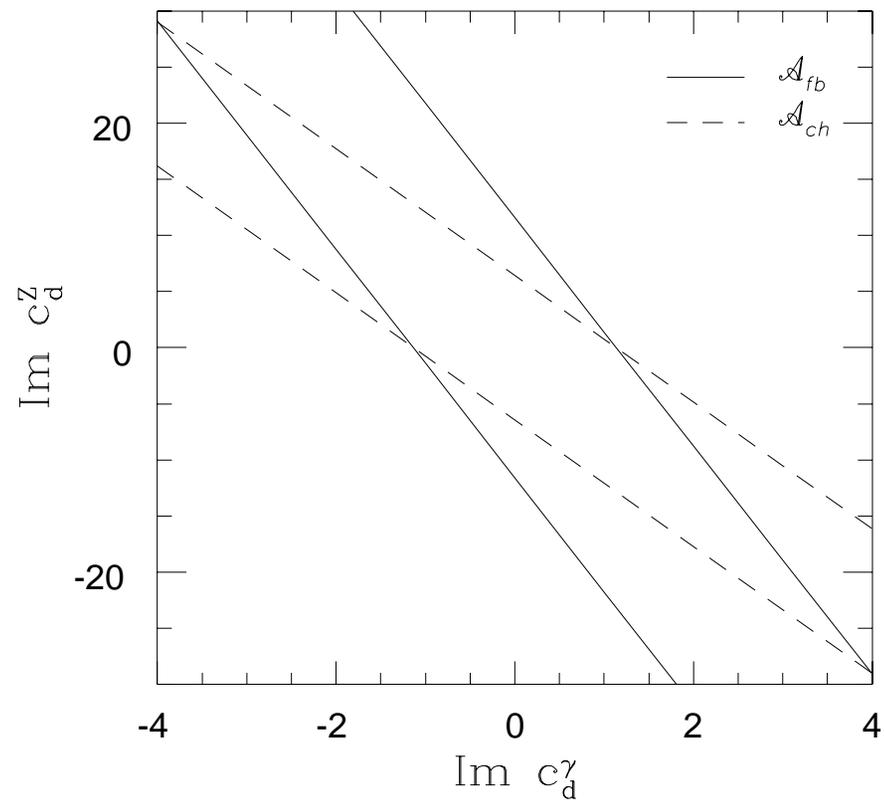
\end{document}